\documentclass[english]{article}
 \pdfoutput=1
\usepackage[T1]{fontenc}
\usepackage[latin9]{inputenc}
\usepackage[a4paper]{geometry}
\usepackage{graphicx}
\usepackage{setspace}
\usepackage{esint}
\makeatletter

\providecommand{\tabularnewline}{\\}

\newcommand{\lyxaddress}[1]{
\par {\raggedright #1
\vspace{1.4em}
\noindent\par}
}

\renewcommand\[{\begin{equation}}
\renewcommand\]{\end{equation}}
\usepackage{times}
\date{}

\makeatother

\usepackage{babel}
\begin{document}

\title{High Power Characterization of Piezoelectric Materials Using the
Burst/Transient Method with Resonance and Antiresonance Analysis}

\author{Husain Shekhani$^{1}$\thanks{Author to whom correspondence should be addressed. Electronic mail:
hns116@psu.edu}, Timo Scholehwar$^{2}$, Eberhard Hennig$^{2}$, and Kenji Uchino$^{1}$}

\maketitle

\lyxaddress{$^{1}$Department of Electrical Engineering, The Pennsylvania State
University, University Park, Pennsylvania, 16801, USA}

\lyxaddress{$^{2}$PI Ceramic GmbH, Lederhose, 07589, Germany}
\begin{abstract}
In this paper, a comprehensive methodology for characterizing the
high power resonance behavior of bulk piezoelectric ceramics using
the burst method is described. In the burst method, the sample is
electrically driven at its resonance frequency, and then either a
short circuit or an open circuit condition is imposed, after which
the vibration decays at the resonance or antiresonance frequency,
respectively. This decay can be used to measure the quality factor
in either of these conditions. The resulting current in the short
circuit vibration condition is related to the vibration velocity through
the \textquotedblleft force factor.\textquotedblright{} The generated
voltage in the open circuit vibration condition corresponds to the
displacement by the \textquotedblleft voltage factor.\textquotedblright{}
The force factor and the voltage factor are related to material properties
and physical dimensions of the sample. Using this method, the high
power behavior of the permittivity, compliance, piezoelectric charge
constant, electromechanical coupling factor, and material losses can
be determined directly by measuring the resonance (short circuit)
and antiresonance (open circuit) frequencies, their corresponding
quality factors, the force factor $A$, and the voltage factor $B$.
The experimental procedure to apply this method is described and demonstrated
on commercially available hard and semi-hard PZT materials of $k_{31}$
geometry. 
\end{abstract}

\section{Introduction}

Piezoelectric materials are used in a variety of high power applications
such as ultrasonic motors and underwater sonar transducers. The properties
of these materials are subject to the conditions in which they are
applied in. Therefore, the properties must be measured in comparable
high power testing environments in order to achieve relevant measurements.
This paper uses the burst/transient method for measuring the properties
of piezoelectric materials in high power conditions using a comprehensive
approach with resonance and antiresonance analysis. The burst method
offers several advantages over other high power measurement methods:
appropriate application of linear theory, short measurement time,
data collection over large range of vibration levels in single measurement,
simplicity in experimental application, and no heat generation/temperature
rise during measurement. 

Two methods are commonly used to determine the material properties
in piezoelectric ceramics in high power resonance conditions: resonance
electrical spectroscopy and the burst/transient method. In resonance
impedance spectroscopy, the sample is continuously driven electrically,
and its electrical impedance or admittance is measured across its
resonance and antiresonance frequency. By fitting an equivalent circuit
or by utilizing a power bandwidth approach, the loss factors and material
properties for a particular vibration mode can be calculated. Using
the resonance frequency, the elastic compliance can be measured. Utilizing
the relative difference between the resonance and antiresonance frequencies,
the electromechanical coupling factor can be calculated. \cite{uchino_loss_2011,hennig2005large}

The second method of measuring properties in high power conditions
is through the burst or transient method. In this method, the ceramic
is driven using a large excitation voltage at its resonance frequency
for a set number of cycles. Then, a short circuit condition is imposed
and the sample's oscillation rings down at its resonance frequency.
In this case, the short circuit current is proportional to the vibration
amplitude. If instead, an open circuit condition is imposed, the sample\textquoteright s
oscillation rings down at its antiresonance frequency. In this case,
the open circuit voltage is proportional to the displacement. By measuring
the rate of signal decay, the loss factors at resonance and antiresonance
can be calculated using an incremental time constant formulation.
Using the resonance and antiresonance frequencies, the elastic compliance
and electromechanical coupling factor can be calculated similarly
to the process of the impedance method. The burst method can be considered
as a mechanical excitation method because electrical stimulation is
not applied during the measurement period. \cite{uchino_loss_2011,umeda_measurement_1998,takahashi_characteristics_1999}.

In the burst/transient method experiment, temperature rise is not
generated due to low driving times (often less than 10ms). Also, data
for a wide range of vibration levels can be obtained from the decaying
oscillation, whereas data must be collected for a single vibration
level at a time using the continuous drive method. It has been shown
that results from the burst method show higher quality factors (lower
losses), and that this is not completely due to temperature rise difference
\cite{uchino_loss_2011,takahashi_characteristics_1999}. However,
to the authors\textquoteright{} knowledge, a rigorous comparison of
these techniques has not been made accounting for temperature distribution
in the ceramic and number drive cycles. This may lead to comparable
results between the two methods. 

The burst method was developed by Umeda et al. for determining the
equivalent circuit parameters of a piezoelectric transducer \cite{umeda_measurement_1998}.
It was thereafter adapted to measure the properties of piezoelectric
ceramic samples \cite{takahashi_characteristics_1999,umeda_effects_1999}.
The analysis has proved useful to many researchers studying the change
of properties with vibration velocity and also to compare property
values between materials, especially between PZT and lead-free materials
\cite{tou_properties_2009,nagata_high_2010,hagiwara_analysis_2011,noumura_high-power_2010,hiruma_piezoelectric_2008}.
However, almost all the analyses have been done at the resonance condition
(short circuit). Chang and coworkers have characterized the response
of a Langevin Transducer at antiresonance by introducing an open circuit
condition to evaluate the equivalent circuit \cite{chang_open-circuit_2003,chang_electrical_2004,chang_investigation_2007}.
Also, the open circuit/antiresonance condition has been applied to
a hard $k_{31}$ PZT material to characterize its antiresonance quality
factor \cite{uchino_loss_2011,ural_high_2010}. However, the analytical
formulation to derive material properties from the voltage factor
and the method to analyze the dielectric permittivity from it will
be discussed in this paper for the first time. 

The objective of this paper is to provide a comprehensive characterization
approach for the $k_{31}$ resonator using the burst method. Some
of the techniques have already been presented by previous researchers,
but this work adds new analytical methods, supporting experimental
techniques, and discussion of physical significance of the high power
measurement results.

Firstly, the derivation of the force factor $A_{31}$ and voltage
factor $B_{31}$ in terms of material properties and sample geometry
will be derived from the constitutive equations for the $k_{31}$
piezoelectric sample. The force factor $A_{31}$ is the relationship
between current and vibration in resonance and it is related to the
effective piezoelectric stress coefficient $e_{31}^{*}$. The force
factor analysis for the $k_{31}$ has been presented previously by
Takahashi \cite{takahashi_characteristics_1999}, but its explicit
derivation has not. The voltage factor $B_{31}$ is the relationship
between open circuit voltage and displacement in resonance, related
with the effective piezoelectric stiffness coefficient $h_{31}^{*}$.
We believe it has not been applied nor analyzed in bulk piezoceramics
in resonance conditions. Additionally, use of the resonance-antiresonance
frequency separation to calculate the electromechanical coupling factor
using the burst method will be presented for the first time. Also,
the theory for measuring the permittivity directly in resonance conditions
will be outlined and demonstrated.

Secondly, the experimental results of the high power resonance response
of a hard PZT and semi-hard PZT material of $k_{31}$ geometry will
be discussed. The quality factors and the various properties will
be measured using application of resonance and antiresonance. The
experimental methods to apply the burst method will be described,
along with supporting data analysis techniques.

\section{Derivation of the force factor and the voltage factor for $k_{31}$
resonators}

In this section, the force factor and the voltage factor for the $k_{31}$
resonator will be developed and the approach for calculating material
properties will be summarized. Also, the method to characterize the
dielectric permittivity in resonance conditions will be described.
Fig.\ 1 shows the geometry of the sample and Tab.\ 1 defines the
symbols used in the derivation as they will appear. The derivations
assume a rectangular plate with $a\ll b\ll L$, fully electroded,
and poled along the thickness ($a$). Therefore, the criteria for
a $k_{31}$ piezoelectric resonator are met. 

The relationship between current and vibration in a resonating piezoelectric
ceramic under short circuit and open circuit conditions will be derived
for the $k_{31}$ mode. These relationships will be used to experimentally
evaluate the behavior of piezoelectric samples tested using the burst
method in the next section. For the discussion in this paper, the
resonance frequency $\omega_{A}$ will be described as A-type resonance
and antiresonance frequency $\omega_{B}$ will be described as B-type
resonance . The $k_{31}$ mode undergoes mechanical resonance during
electrical resonance, corresponding to $s_{11}^{E}$. This mode undergoes
an electromechanical coupling resonance at its antiresonance frequency,
effectively electrically coupling the motional and damped branch of
the resonator (Fig.\ 2) \cite{ikeda_fundamentals_1996}. Electrical antiresonance
is achieved by an open circuit ($D$ constant) condition and electrical
resonance is achieved by a short circuit condition ($E$ constant).

As long as the sample is symmetric, both in its geometry and boundary
conditions, the mode shape will be symmetric about the center of the
sample. In general, the mode shape of a piezoelectric resonator with
stress free boundary conditions, undergoing vibration in one dimension,
with losses, and having finite displacement can be described as

\[
u(x,t)=u_{0}f(x)\sin(\omega t),
\]
where $f(x)$ is a function symmetric about the origin normalized
to the displacement at the ends of the piezoelectric resonator, where
$f(0)=0$. Strain is defined as

\begin{equation}
\partial u/\partial x=u_{0}f'(x).
\end{equation}
Then, according to the fundamental theorem of calculus

\begin{equation}
\int_{-L/2}^{L/2}\frac{\partial u}{\partial x}\,\mathrm{dx}=u(L/2,t)-u(-L/2,t)=2u_{0}\sin(\omega t).\label{eq:fundamental}
\end{equation}

In general, the vibration or displacement distribution (mode shape)
for a $k_{31}$ resonator and the $k_{33}$ resonator is sinusoidal.
In general, its mode shape as a function of frequency and normalized
to the edge displacement is : for the $k_{31}$ mode \cite{ikeda_fundamentals_1996}

\begin{equation}
u(x)=u_{0}\frac{\sin\Omega_{31}x}{\sin\Omega_{31}L/2}.\label{eq:k31 mode shape}
\end{equation}
where $\Omega_{31}=\omega/\nu_{11}^{E}$ and $\nu$ is the speed of
sound in the material $\nu_{11}^{E}=1/\sqrt{\rho s_{11}^{E}}$. These
mode shape functions define the displacement distribution at resonance
modes (e.g.\ 1st, 2nd, and 3rd mode) and also at frequencies in between.
The normalized mode shapes at the first three mechanical resonance
modes of the $k_{33}$ and $k_{31}$ resonators are given in Fig.\ 3.
The displacement distribution of frequencies in between resonance
modes have distributions of partial wavelengths, unlike mechanical
resonance modes, whose distributions of wavelengths of $\frac{1}{2}+(n-1)$
of the geometry in the direction of vibration. 

According to Eq.\ (\ref{eq:fundamental}), Eqs.\ (\ref{eq:k31 mode shape})
can be written as

\begin{equation}
\int_{-L/2}^{L/2}\frac{\partial\left(u_{0}\frac{\sin\Omega_{31}x}{\sin\Omega_{31}L/2}\right)}{\partial x}\,\mathrm{dx}=2u_{0}.\label{eq:simple k31 strain}
\end{equation}
Eq.~(\ref{eq:simple k31 strain}) does not assume a particular frequency,
so the results are true at the mechanical resonance frequency and
at other frequencies. Depending on the mode type, either the electrical
antiresonance frequency or electrical resonance frequency is the mechanical
resonance frequency determined by the speed of sound in the direction
of vibration. However, because the derivation is general, the specific
vibration distribution in question will not need to be explicitly
handled in the derivation because they have been solved in the general
case in Eq. (\ref{eq:simple k31 strain}).

\subsection{Short circuit analysis of the force factor}

The constitutive equation describing the electric displacement of
a piezoelectric $k_{31}$ resonator is \cite{ikeda_fundamentals_1996}

\begin{equation}
D_{3}(t)=d_{31}X_{1}+\varepsilon_{33}^{X}\varepsilon_{0}E_{3}(t).\label{eq:constitutive-2}
\end{equation}
By using the electromechanical coupling factor, this equation can
be rewritten as

\begin{equation}
D_{3}(t)=e_{31}^{*}\frac{\partial u}{\partial x}+\varepsilon_{33}^{x_{1}}\varepsilon_{0}E_{3}(t),\label{eq:constitutive}
\end{equation}
where $e_{31}^{*}$ is the effective piezoelectric stress coefficient
defined as $e_{31}^{*}=d_{31}/s_{11}^{E}$ and $\varepsilon_{33}^{x_{1}}$
is the relative permittivity having clamping in the length (1-direction).
The actual piezoelectric stress coefficient $e_{31}$(non-star) is
defined from clamped boundary conditions in the 2-direction and the
3-direciton, whereas for the $k_{31}$ resonator these boundaries
are stress free. Thus, the effective piezoelectric stress coefficient
is used as defined above. For the electrical boundary condition of
zero electric potential case (short circuit), the external field is
equal to zero. Therefore,

\begin{equation}
D_{3}(t)=e_{31}^{*}\frac{\partial u}{\partial x},
\end{equation}
and
\[
\dot{D}_{3}=e_{31}^{*}\frac{\partial^{2}u}{\partial x\partial t}.
\]
The current can be written as

\[
i(t)=\int_{A_{e}}\dot{D}_{3}\mathrm{\,dA_{e}},
\]
where

\[
\mathrm{dA_{e}}=b\,\mathrm{dx}.
\]
Therefore,

\[
i(t)=b\int_{-L/2}^{L/2}\dot{D}_{3}\,\mathrm{dx}.
\]
Assuming that the sample is undergoing free vibration at the resonance
frequency (constant E/short circuit conditions), we can apply Eq.\
(\ref{eq:fundamental}) such that the current can be written as

\[
i_{0}=-2e_{31}^{*}u_{0}b\omega_{A}.
\]
This equation can also be written in terms of vibration velocity at
the plate edge ($x=\pm L/2$), given $\omega u_{0}=v_{0}$ for sinusoidal
time varying displacement

\begin{equation}
i_{0}=-2e_{31}^{*}bv_{0}.
\end{equation}

The force factor $A_{31}$, defined as the ratio between short circuit
current and edge vibration velocity, can then be written as

\begin{equation}
A_{31}=\frac{i_{0}}{v_{0}}=-2e_{31}^{*}b=-2b\frac{d_{31}}{s_{11}^{E}}.\label{eq:force factor}
\end{equation}

\subsection{Open circuit analysis of the voltage factor}

For open circuit conditions the total electric displacement is equal
to zero

\[
\int_{-L/2}^{L/2}D_{3}(t)\,\mathrm{dx_{1}}=0.
\]
Therefore, the constitutive equation described in Eq.\ (\ref{eq:constitutive})
can be written as

\begin{equation}
\int_{-L/2}^{L/2}D_{3}(t)\,\mathrm{dx_{1}}=\int_{-L/2}^{L/2}\left(e_{31}^{*}\frac{\partial u}{\partial x}+\varepsilon_{33}^{x_{1}}\varepsilon_{0}E_{3}(t)\right)\,\mathrm{dx_{1}}.\label{eq:constitutive-1}
\end{equation}
Assuming the variation of strain in thickness is negligible, the electric
field across the thickness is uniform. Therefore, the $E_{3}(x,t)=-V(t)/a$.
Integrating across the length of the resonator,
\begin{equation}
\int_{-L/2}^{L/2}E_{3}(x,t)\,\mathrm{dx}=-\int_{-L/2}^{L/2}V(t)/a\,\mathrm{dx}=\frac{e_{31}^{*}}{\varepsilon_{33}^{x_{1}}\varepsilon_{0}}\int_{-L/2}^{L/2}\frac{\partial u}{\partial x}\,\mathrm{dx}.
\end{equation}
This equation can be rewritten using Eq.\ (\ref{eq:fundamental})
assuming natural vibration at the antiresonance frequency in open
circuit conditions

\[
LV_{0}/a=\frac{e_{31}^{*}}{\varepsilon_{33}^{x_{1}}\varepsilon_{0}}2u_{0}.
\]
Thus, the relationship between displacement and generated open circuit
voltage for a $k_{31}$ resonator in its antiresonance mode can be
written as

\[
V_{0}=\frac{2ae_{31}^{*}}{L\varepsilon_{33}^{x_{1}}\varepsilon_{0}}u_{0},
\]
and the voltage factor ($B_{31}$), the ratio between open circuit
voltage and edge displacement, can be written as

\begin{equation}
B_{31}=\frac{V_{0}}{u_{0}}=\frac{2a}{L}\frac{e_{31}^{*}}{\varepsilon_{33}^{x_{1}}\varepsilon_{0}}=\frac{2a}{L}\frac{g_{31}}{s_{11}^{D}}=\frac{2a}{L}h_{31}^{*},\label{eq:B}
\end{equation}
given negligible cross coupling. The effective piezoelectric stiffness
coefficient $h_{31}^{*}$ the electric field generated in the 3 direction
(polarization direction) for an applied strain in the 1 direction
under constant $D$ conditions (open circuit) having free boundary
conditions in the 2-direciton and the 3-direction:

\[
E_{3}=h_{31}^{*}\frac{\partial u_{1}}{\partial x_{1}},
\]
 the relationship between stress and strain under constant $D$

\[
x_{1}=s_{11}^{D}X_{1},
\]
and the relationship between generated electric field under constant
$D$ and stress $X$ is 

\[
E_{3}=g_{31}X_{1}.
\]

\subsection{Analysis of elastic compliance, piezoelectric charge coefficient,
and the electromechanical coupling factor }

It is well known that the sound velocity in a $k_{31}$ resonator
propagating in the 1-direction occurs is governed by the density and
the elastic compliance under constant electric field \cite{ikeda_fundamentals_1996}.
The first resonance frequency in the $k_{31}$ resonator corresponds
to the $s_{11}^{E}$ according to the equation 

\[
s_{11}^{E}=1/((2Lf_{A})^{2}\rho).
\]
By utilizing the measurement of the force factor, the piezoelectric
charge coefficient can be computed

\[
d_{31}=-A_{31}s_{11}^{E}/2b.
\]
A more common approach to calculate this coefficient, frequently used
in electrical resonance spectroscopy, is as follows: the off-resonance
permittivity and resonance elastic compliance can be used to separate
the piezoelectric charge coefficient from the coupling coefficient.
$k_{31}$ is calculated from a trigonometric function whose variables
are the resonance and antiresonance frequencies \cite{ikeda_fundamentals_1996}.
Therefore, $d_{31}$ can be expressed as 

\[
d_{31}=-k_{31}\sqrt{s_{11}^{E}\varepsilon_{33}^{X}\varepsilon_{0}}.
\]
This approach assumes that the $\varepsilon_{33}^{X}$ does not change
in resonance conditions. The calculation of $d_{31}$ using the force
factor does not make this assumption, so it is expected to be more
accurate. A detailed experimental analysis will follow in the next
section. 

By using the piezoelectric stress coefficient calculated at resonance
(from the force factor) and the converse piezoelectric constant calculated
at antiresonance (from the newly derived voltage factor), the clamped
permittivity can be calculated in resonance conditions directly. Then,
$\varepsilon_{33}^{X}$ can be calculated using the $k_{31}^{2}$. 

\begin{equation}
\varepsilon_{0}\varepsilon_{33}^{X}(1-k_{31}^{2})=\varepsilon_{0}\varepsilon_{33}^{x_{1}}=\frac{e_{31}^{*}}{h_{31}^{*}}=\frac{A_{31}}{B_{31}}\frac{a}{Lb}\label{eq:perm final}
\end{equation}
Permittivity has never been calculated directly in resonance conditions
according to the authors\textquoteright{} knowledge. Takahashi et
al. has reported permittivity in resonance conditions, but assumes
that only the motional capacitance changes and the clamped capacitance
in resonance does not change. The validity of this assumption will
be evaluated from the data in the next section.

\section{Excitation of resonance and antiresonance using burst drive}

In the burst mode experiment, the sample is first driven with an oscillating
voltage in order to build up vibration. Then, the driving signal is
removed, and thus the sample vibration decays at the system's natural
frequency. Fig.\ 4a demonstrates the application of the burst mode
using resonance methods (constant $E$). The sample is driven at its
resonance frequency for a set number of cycles, after which a short
circuit condition is imposed, inducing natural vibration at the electrical
resonance frequency. Fig.\ 4b demonstrates the case where an open
circuit condition is imposed, generating antiresonance vibration.
For the open circuit condition, a bias voltage (not shown) can be
generated along with the decaying oscillating voltage as reported
by Chang \cite{chang_open-circuit_2003}. This bias voltage is generated
from the charge on the electrode at the time of introducing the open
circuit condition. Depending on the instantaneous charge before the
open circuit condition is imposed, the bias voltage can be positive,
negative, or zero. By driving the sample near the antiresonance frequency
before the open circuit is introduced, the resulting bias voltage
is greatly reduced because of the small current in this condition.
This was the unique approach taken in this research for measuring
the antiresonance burst response. 

By applying the burst mode at resonance (short circuit) and antiresonance
(open circuit) conditions, the loss factors and the real properties
of the material can be measured. For a damped linear system oscillating
at its natural frequency, the quality factor can be described using
the relative rate of decay of vibration amplitude. In general, \cite{umeda_measurement_1998}

\begin{equation}
Q=\frac{2\pi f}{2\ln(\frac{v_{1}}{v_{2}})/(t_{2}-t_{1})}.
\end{equation}
This equation is true at both resonance and antiresonance. At resonance,
the current is proportional to the vibration velocity; therefore,
its decay can be used. Similarly, the voltage decay can be used at
antiresonance to determine the quality factor at antiresonance.

\section{Experimental application}

\subsection{Sample selection}

The burst mode experiment was performed on $k_{31}$ samples of commercially
available PZT compositions: three PIC 184 (PI Ceramic, Germany) $k_{31}$
samples and three PIC 144 (PI Ceramic, Germany) $k_{31}$ samples,
whose low power properties are listed in Tab.\ 1. All the samples
have a geometry of $\mathrm{40\times6\times1mm}$$^{3}$. Commercially
available materials were chosen because they provide the most relevant
results from a practical perspective. The samples were supported from
the center (nodal point) by adjustable spring loaded electrodes. Each
sample in each condition (resonance and antiresonance) was measured
twice. After every measurement, the sample was removed, turned over,
and reloaded into the sample holder. Therefore, each data point presented
is an average of six measurements and the error bars are the standard
deviation of these measurements. Previous researchers either did not
present multiple measurements or did not discuss the effect of reloading
the sample on the standard deviation \cite{umeda_measurement_1998,umeda_effects_1999,takahashi_nonlinear_2000}.
By measuring one sample several times without removing and reloading
it, the standard deviation is significantly reduced, but the true
value of the material property remains more ambiguous. Therefore,
the fact that the samples were removed and reloaded in the sample
holder is highlighted. The measurement of the properties of piezoelectric
material having low losses, such as hard PZT, are more sensitive to
the sample holder positioning.

\subsection{Experimental setup and procedure}

The experimental setup diagram can be seen in Fig.\ 5. The sample
was driven by a function generator (Siglent SDO342) through a power
amplifier (NF4010). The sample was held at the nodal point. The drive
period was set at 10 cycles. For the short circuit measurement, the
sample was driven at its resonance frequency at 50$\mathrm{V/mm}$
using the burst mode of the function generator. After the burst signal
finished, the applied voltage to the sample was 0$\mathrm{V}$, which
is equivalent to a short circuit condition. The current and vibration
then decayed. The measured voltage during the short circuit condition
indicated an added effective inductive load on the order of 1$\Omega$.
This was due to the current flow through the BNC cable from the amplifier.
After applying the burst signal, antiresonance vibration was induced
by isolating the sample from the driving signal using an electromechanical
relay. The electromechanical relay was triggered with a precise time
delay from the burst signal using the second channel of the function
generator.

For the open circuit experiments, the sample was driven close to its
antiresonance frequency (minimum current), and for the short circuit
experiments, the sample was driven close to its resonance frequency
(maximum current). This is because after imposing the open circuit
or short condition, the sample\textquoteright s oscillating frequency
immediately shifts to the frequency particular to the electrical boundary
condition, antiresonance or resonance, respectively. If this frequency
shift is significant, transient higher harmonics are generated and
they cause the vibration decay to become irregular and difficult to
analyze. By driving the system at its resonance frequency prior to
applying the short circuit and the antiresonance frequency prior to
applying open circuit, transient higher harmonics were minimized and
random error was reduced significantly.

The current was measured by a 10x wire loop and a current probe (Tektronix
TCPA300 w/ TCP 305 probe). The current probe was verified to produce
accurate readings for the current levels tested by measuring the current
produced in a 1$\mathrm{k}\Omega$ precision resistor near the experimental
resonance frequency.

For the open circuit experiments, a electromechanical relay was used
to isolate the sample from the signal. The function generator used
has two channels, allowing for a programmable delay between them.
One of the channels was responsible for providing the burst signal
(10 sinusoidal cycles) through the amplifier, and the other channel
was responsible for driving the relay directly. The channel powering
the relay was normally on at 10V, and it decreased to 0V after the
trigger was initiated. The delay for the turn off time of the relay
was approximately 0.45ms. This delay time was added to the burst signal
channel along with the delay time needed to allow for the 10 drive
cycles to the sample to build up vibration. The open circuit voltage
was measured using a 100x voltage probe (Tektronix P1500). The voltage
probe had a resistance of 10M$\Omega$, much higher than the antiresonance
impedance of the samples, and the capacitance of the probe was more
than a thousand times less than that of the samples measured. 

Additional measurement issues should be taken into consideration when
selecting the sample geometry, in addition to isolating the vibration
mode of analysis. A sufficient width for the sample should be chosen
to generate the level of current needed according to the force factor
for the equipment to make accurate measurements. Similarly, for antiresonance,
the $a/L$ ratio changes the voltage factor. Usually, voltage can
be measured with high accuracy using an oscilloscope without special
considerations. Depending on the thickness and length, over a 400$\mathrm{V}$
can be generated, as reported by Uchino \cite{uchino_loss_2011},
so therefore the $a/L$ ratio may need to be considered for the sample
fabrication in order to avoid voltage levels which may damage measurement
equipment.

\subsection{Data collection method}

A crucial factor in the attainment of repeatable and accurate results
was the data collection and analysis method used, and the storage
capacity of the oscilloscope. A LabVIEW program utilizing a function
which estimates the frequency and amplitude of the dominant harmonic
signal from a Fourier transform analysis was applied. It was found
that the built-in FFT analysis function of the oscilloscope lacks
sufficient resolution, both in frequency and amplitude measurements,
to provide systematic results; analyzing the waveforms on the PC provided
better final results. Small instability in the signal can cause large
error in frequency measured from zero crossing of the oscillating
signals and amplitude characterization using maximum values. Therefore,
a FFT analysis is the most appropriate. The name of the LabVIEW function
used to accomplish this was called \textquotedblleft Extract Single
Tone,\textquotedblright{} by which the frequency and amplitude of
the signal were measured by interpolating the FFT spectrum of the
waveform. The oscilloscope time base was initially set to be large
enough to allow the vibration velocity to decrease ten times from
the maximum value within a single capture of the oscilloscope. After
this acquisition, a smaller time base was used to approximate the
capture of data at specific instances in time. By moving the viewing
window forward (with the small time base), data for different vibration
levels could be measured. For the hard PZT measured in this study,
the smaller time base used was 4 cycles, and for the semi-hard PZT
the time base used was 2 cycles. A smaller number of cycles must be
used for the semi-hard PZT because its vibration signal decays more
rapidly. The initial use of the large time base for the waveform acquisition
and then the subsequent decrease in the time base showed no apparent
difference in measurement from using a small time base for the initial
acquisition. The ability of the FFT function used over several cycles
increased the measurement integrity significantly over using the peak
values of the waveforms to determine signal amplitude, decay, and
frequency.

\section{Results and discussion}

\subsection{Measurement of the force factor and voltage factor}

Regarding resonance characterization, the ratio between the short
circuit current and edge vibration velocity was used to calculate
the force factor and the piezoelectric stress constant, $e_{31}^{*}=d_{31}/s_{11}^{E}$,
according to Eq.~(14). Using the resonance frequency, the compliance
was calculated, and the piezoelectric charge coefficient as well using
the piezoelectric stress coefficient. The resonance characterization
for PIC 144 and PIC 184 is shown in Figs.~6a and 6b. PIC 144 shows
lower property values ($d_{31}$, $s_{11}^{E}$, $e_{31}^{*}$) than
PIC 184, which is expected because PIC 144 is a hard PZT composition
and has a higher quality factor judging from its low power properties
(Tab.\ 2). The properties $d_{31}$, $s_{11}^{E}$, and $e_{31}^{*}$
of PIC 184 and PIC 144 change linearly with vibration velocity. However,
PIC 144 has more stable properties with vibration velocity but a larger
deviation from a linear change in material properties with vibration
velocity. That being said, the change in properties of PIC 144 with
respect to the low power properties (lowest vibration level) is significantly
less than that of PIC 184. 

By utilizing the displacement and open circuit voltage at antiresonance,
the voltage factor $B_{31}$ and the effective piezoelectric stiffness
coefficient $h_{31}^{*}$ were calculated using Eq.~(20) (Fig.\
7). The coupling factor was calculated using the resonance and antiresonance
frequencies, and the trigonometric function defined for the $k_{31}$
resonator \cite{ikeda_fundamentals_1996}.
The coupling factor increases with the vibration velocity; the change
in the coupling factor with vibration velocity (slope) of PIC 184
is three times larger than that of PIC 144. $h_{31}^{*}$ decreases
with increasing vibration velocity, contrary to the trend of the other
properties. That being said, it has a much smaller dependence on vibration
velocity than the other properties, namely those determined at resonance.

\subsection{Characterization of losses at resonance and antiresonance }

Using the decay of vibration at resonance and antiresonance, the quality
factors were calculated. Each data point used amplitude data from
two vibration measurements; therefore, the scale was readjusted as
an average of the vibration velocity. Fig.\ 8 shows the results;
a log-log plot was used to easily distinguish and compare the trends
between the two compositions. PIC 144 shows stable characteristics
of the quality factor, until about 150 mm/s RMS, after which a sharp
degradation in the quality factors occurred. PIC 184, however, showed
an immediate decrease in its quality factors. $Q_{B}$ was larger
than $Q_{A}$ for both the materials.

\subsection{Characterization of permittivity $\varepsilon_{33}^{X}$ in high
power resonance conditions}

\subsubsection{Off resonance approximation of the permittivity }

Traditionally, the dielectric permittivity is measured by applying
an oscillating electric potential at an off-resonance frequency and
measuring the resulting charge or current \cite{ikeda_fundamentals_1996}. The dielectric
permittivity has not been measured at resonance high power conditions
because the dielectric response in this condition does not display
a distinct characteristic which can be measured to compute it. Therefore,
researchers have used one of the two approaches to estimate the permittivity
in high power conditions:
\begin{enumerate}
\item Assume the permittivity measured in off-resonance conditions applies
to resonance conditions. This approach is problematic because the
stress conditions and the frequency is different at resonance, and
therefore the property is expected to change, similar to other properties.
\item The other approach is to assume that the permittivity in high power
conditions can be considered as a perturbation of the off-resonance
permittivity using a variation in the motional capacitance, which
is proportional to $d_{31}^{2}/s_{11}^{E}$ for the $k_{31}$ mode.
\end{enumerate}
The first approach is flawed because the elastic and piezoelectric
properties of the material are known to change with applied stress;
therefore, the dielectric response must also follow a similar tendency
because $s^{E},d,$ and $\varepsilon^{X}$ all have extrinsic contributions
from non-$180^{\circ}$ domain walls which are strongly affected by
applied stress. Researchers using the second approach assume that
clamped permittivity $\varepsilon^{x}$ of the material is the same
in high power resonance conditions and in low power off-resonance
conditions. The permittivity under constant stress ($\varepsilon_{33}^{X}$)
is a combination of the motional and clamped dielectric response of
the material. Assuming $\varepsilon^{x}$ from off-resonance measurements,
adjusting the motional contribution to $\varepsilon^{X}$ according
the piezoelectric charge constant, the compliance can determine the
permittivity $\varepsilon^{X}$ in these conditions. The basis for
the assumption of the equivalency of the lower power and high power
clamped permittivity is as follows:
\begin{quotation}
The dielectric permittivity $\varepsilon$ in PZT thin films are much
lower than the permittivity $\varepsilon^{X}$ in bulk PZT. This is
because the substrate effectively clamps the PZT film, and thereby
significantly reduces the non-$180^{\circ}$ domain wall motion and
its contribution to the permittivity. The $180^{\circ}$ domain wall
motion contribution to the permittivity, however, is relatively undisturbed
because it does not result in change in strain and therefore is relatively
unaffected by the substrate~\cite{xu2001domain}. In the $k_{31}$
resonator examined, $\varepsilon^{x}$ refers to the dielectric response
of the material with an applied field in the 3-direction and polarization
in the 3-direction with clamping in the 1-direction. This clamping
results is far less clamping than the substrate clamping in PZT thin
films. Because the effective stresses applied due to resonance vibration
of the $k_{31}$ resonator are in the 1-direction, and the $\varepsilon^{x}$
represents the permittivity after ``clamping out'' the domain wall
contributing to domain wall motion the 1-direction, the following
equality should be true: $\varepsilon_{low\,power}^{x}=\varepsilon_{high\,power}^{x}$. 
\end{quotation}
This was the approach taken by Takahashi et.\ al. They have reported
permittivity in resonance conditions, but assume that only the motional
capacitance changes and the clamped capacitance in resonance does
not \cite{takahashi_characteristics_1999}. For the burst method,
they have reported the motional capacitance ($d_{31}^{2}/s_{11}^{E}$)
to remain constant, and therefore the permittivity, which is calculated
partially from off-resonance measurements, to be constant as well.

That being said, there may exist other phenomena which also affect
the clamped permittivity. This may include the frequency response.
The clamped permittivity may change with increasing frequency due
to the dependence of domain wall motion spectral response. Therefore,
this assumption may be invalid for this or other reasons, and it is
preferred to calculate the permittivity using only data from resonance
measurements in order to ensure compatibility between measurements
to resolve a final property.

\subsubsection{Determination of the permittivity using the voltage factor and force
factor in resonance conditions }

By using the piezoelectric stress coefficient calculated at resonance
(from the force factor) and the converse piezoelectric constant calculated
at antiresonance (from the newly derived voltage factor), the clamped
permittivity can be calculated in resonance conditions directly. Then
from Eq.~\ref{eq:perm final-1} $\varepsilon_{33}^{X}$ can be calculated
using the $k_{31}^{2}$:

\begin{equation}
\varepsilon_{0}\varepsilon_{33}^{X}(1-k_{31}^{2})=\varepsilon_{0}\varepsilon_{33}^{x_{1}}=\frac{e_{31}^{*}}{h_{31}^{*}}=\frac{A_{31}}{B_{31}}\frac{a}{Lb}.\label{eq:perm final-1}
\end{equation}

The change in permittivity with vibration velocity can be seen on
Fig.\ 9. The off-resonance permittivity measured for the samples
is in good agreement with the low vibration velocity permittivity
measured through the burst technique as seen in this figure. The off-resonance
permittivity is represented by a star symbol. The clamped and free
permittivity are both changing with increasing vibration velocity.
The permittivity of PIC 184 is larger than that of PIC 144, and this
is to be expected because PIC 184 is a semi-hard PZT with larger off-resonance
permittivity. From the low vibration state to the high one, the permittivity
of both compositions increase. However, the increase in PIC 184 is
larger, demonstrating that its properties have a larger dependence
on vibration conditions. 

As mentioned earlier in this section, we expect $\varepsilon_{low\,power}^{x}=\varepsilon_{high\,power}^{x}$
because of the clamping out of non-$180^{\circ}$ domain walls. However,
the result shown in this study demonstrates that a majority of the
change seen in the free permittivity can actually be attributed to
the clamped permittivity change. The open circuit voltage generated
during antiresonance may have caused the property changes being reported
in this study. For the PIC 184 samples, the largest voltage generated
was 85$\mathrm{V/mm\,RMS}$ @ 450$\mathrm{mm/s\,RMS}$, and for the
PIC 144 samples the largest voltage was 120$\mathrm{V/mm\,RMS}$ @
600$\mathrm{mm/s\,RMS}$. The open circuit voltage generated may cause
domain reorientation in the 180$^{\mathrm{o}}$ domain wall regions.
The motional capacitance (proportional to the difference, $\varepsilon_{33}^{X}-\varepsilon_{33}^{x_{1}}$)
remains fairly constant, so it cannot explain the change in permittivity.

The difference in the $\varepsilon_{low\,power}^{x}$ and $\varepsilon_{high\,power}^{x}$
may also be due to the details of the clamping applied. As mentioned
earlier, in the $k_{31}$ resonator examined, the clamping considered
is the 1-direction only. Therefore, it should not be expected that
all of non-$180^{\circ}$ domain walls are clamped because the clamping
is dissimilar from the substrate clamping effect seen in thin films.
From the data in Fig.~9, it can be said that the change in clamped
permittivity represents the stresses in the 1-direction affects non-$180^{\circ}$
domain wall motion in the 2 and 3 direction.

\section{Summary}

Traditionally, characterization of piezoelectric materials in resonance
is accomplished using continuous drive methods with impedance spectroscopy.
However, the burst/transient method can be used instead. It affords
simple experimental application, reasonable application of linear
theories, and naturally lends itself toward high power measurements.
In this chapter, a comprehensive measurement approach toward characterization
of the $k_{31}$ resonator was demonstrated using resonance and antiresonance
drive methods. The experimental procedure to achieve both of these
conditions was described. The measurement incorporated the use of
the force factor analysis to characterize the elastic compliance and
piezoelectric charge coefficient as used by other researchers. Antiresonance
analysis using the voltage factor was performed for the first time.
In antiresonance, the ratio between open circuit voltage and displacement
gives the voltage factor, which is proportional to the converse piezoelectric
coefficient. With increasing vibration level, the elastic compliance
(constant electric field), piezoelectric charge constant, and piezoelectric
stress constant increased. At the same time, the converse piezoelectric
constant and the quality factors decreased. Also, the calculation
of the permittivity in resonance conditions only using measurements
from resonance conditions was demonstrated. The results show that
the change in the clamped and free permittivity for the $k_{31}$
resonators tested were of similar levels; therefore, the change in
motional capacitance was small and change in clamped capacitance dominated
the dependence of $\varepsilon_{33}^{X}$ on vibration level.

\subsubsection*{Acknowledgment}

The authors would also like to acknowledge the Office of Naval Research
for sponsoring this research under grant number: ONR N00014-12-1-1044.

\bibliographystyle{unsrt}
\bibliography{burstbib2016}

\pagebreak{}

\section*{Figures}

\begin{minipage}[t]{1\columnwidth}%
\includegraphics[width=10cm]{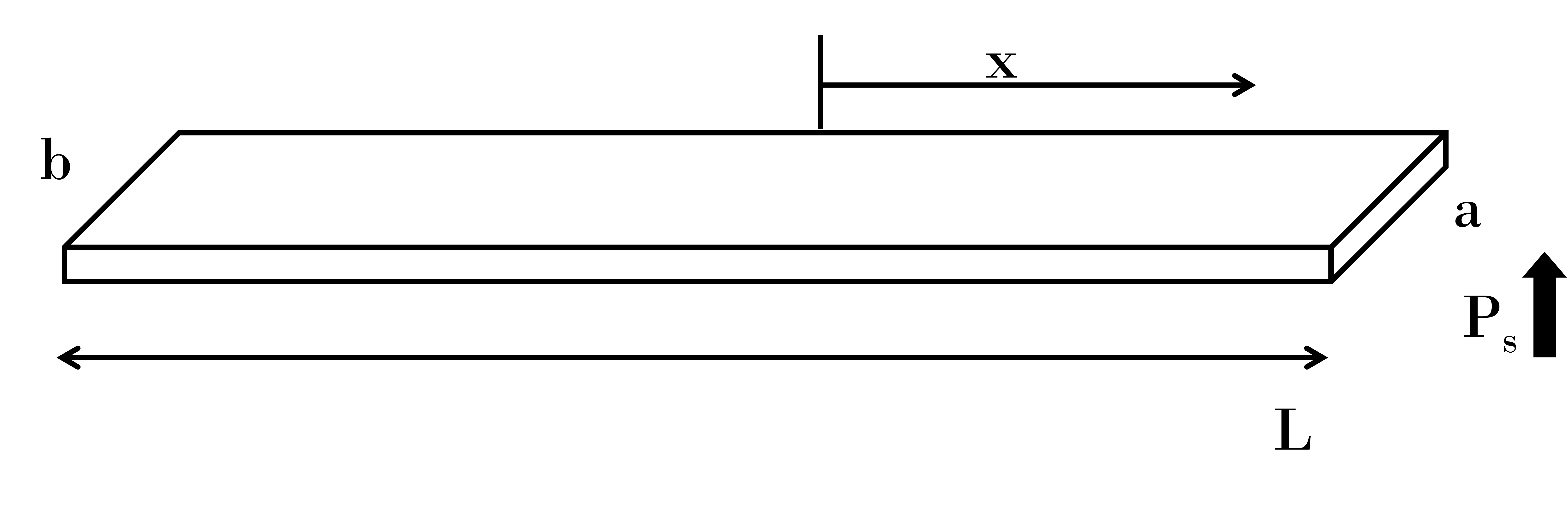}
\begin{description}
\item [{Figure~1}] Geometry of a $k_{31}$ resonator\end{description}
\end{minipage}

\begin{minipage}[t]{1\columnwidth}%
\includegraphics[width=10cm]{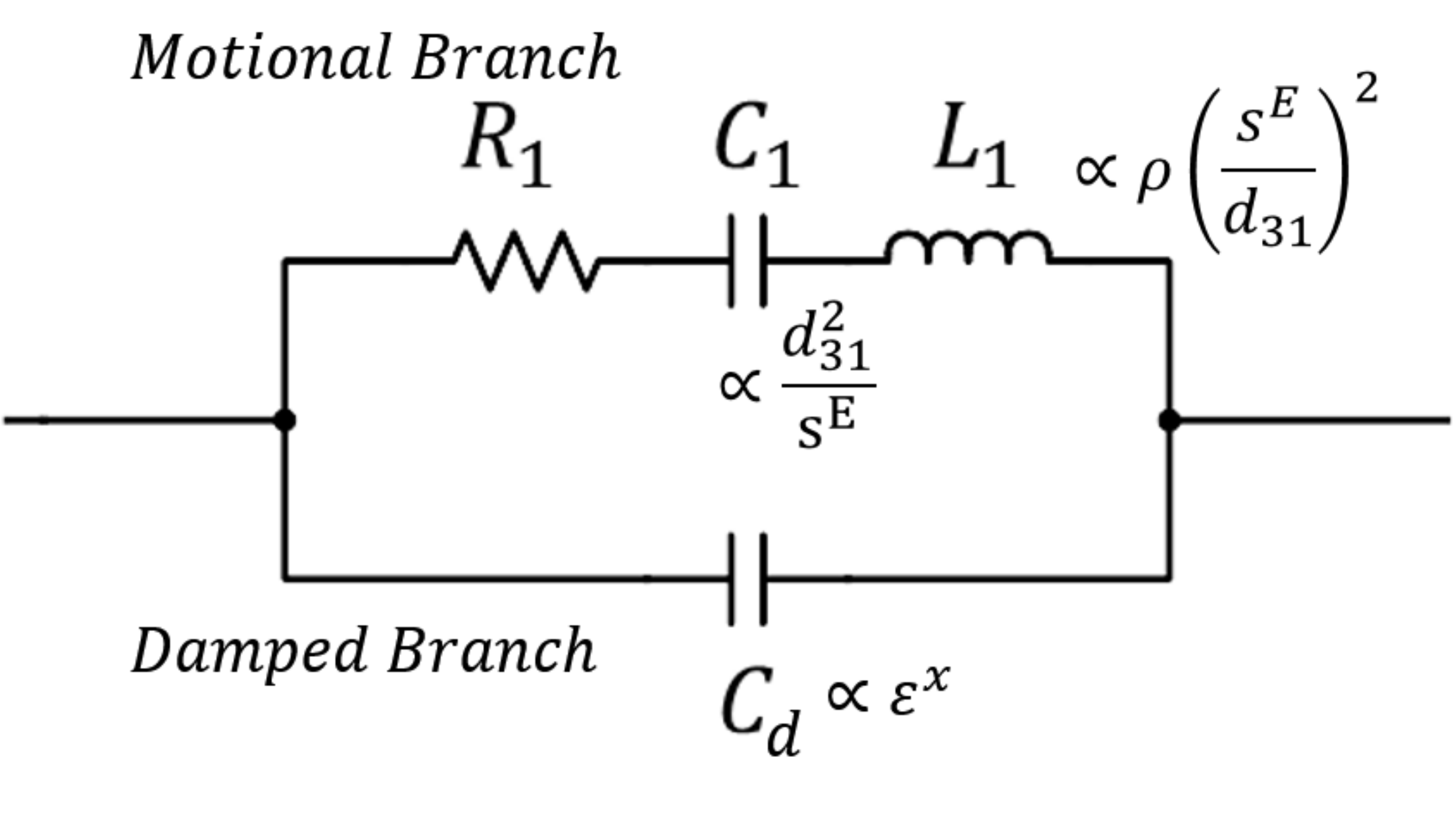}
\begin{description}
\item [{Figure~2}] Simplified equivalent circuit of a $k_{31}$ piezoelectric
resonator (material property equivalencies are presented for illustration)\end{description}
\end{minipage}

\begin{minipage}[t]{1\columnwidth}%
\includegraphics[width=10cm]{long-mode-shape}
\begin{description}
\item [{Figure~3}] Displacement distribution for the first three resonance
modes for the $k_{31}$ resonator\end{description}
\end{minipage}

\begin{minipage}[t]{1\columnwidth}%
\includegraphics[width=7cm]{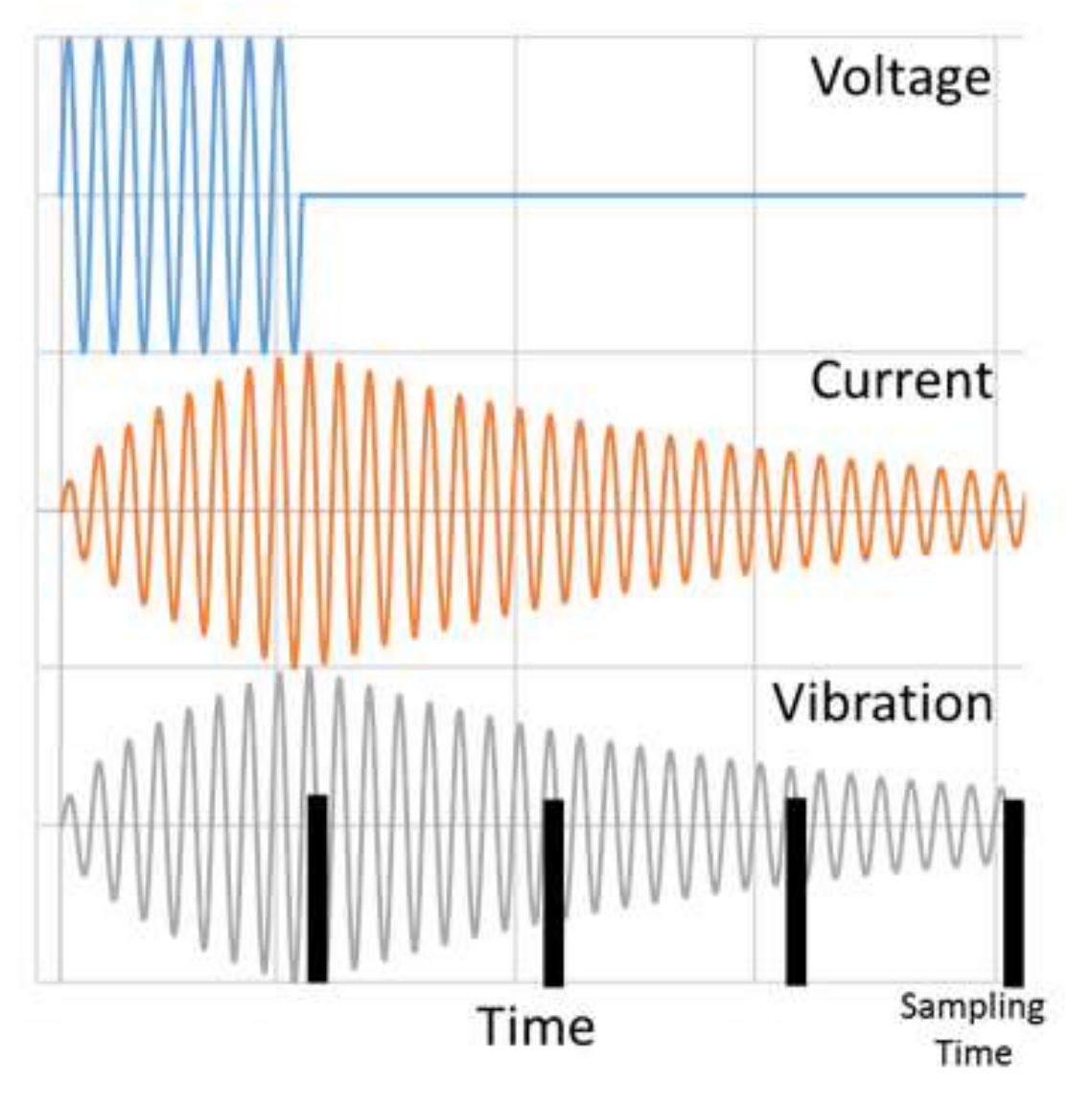}\includegraphics[width=7cm]{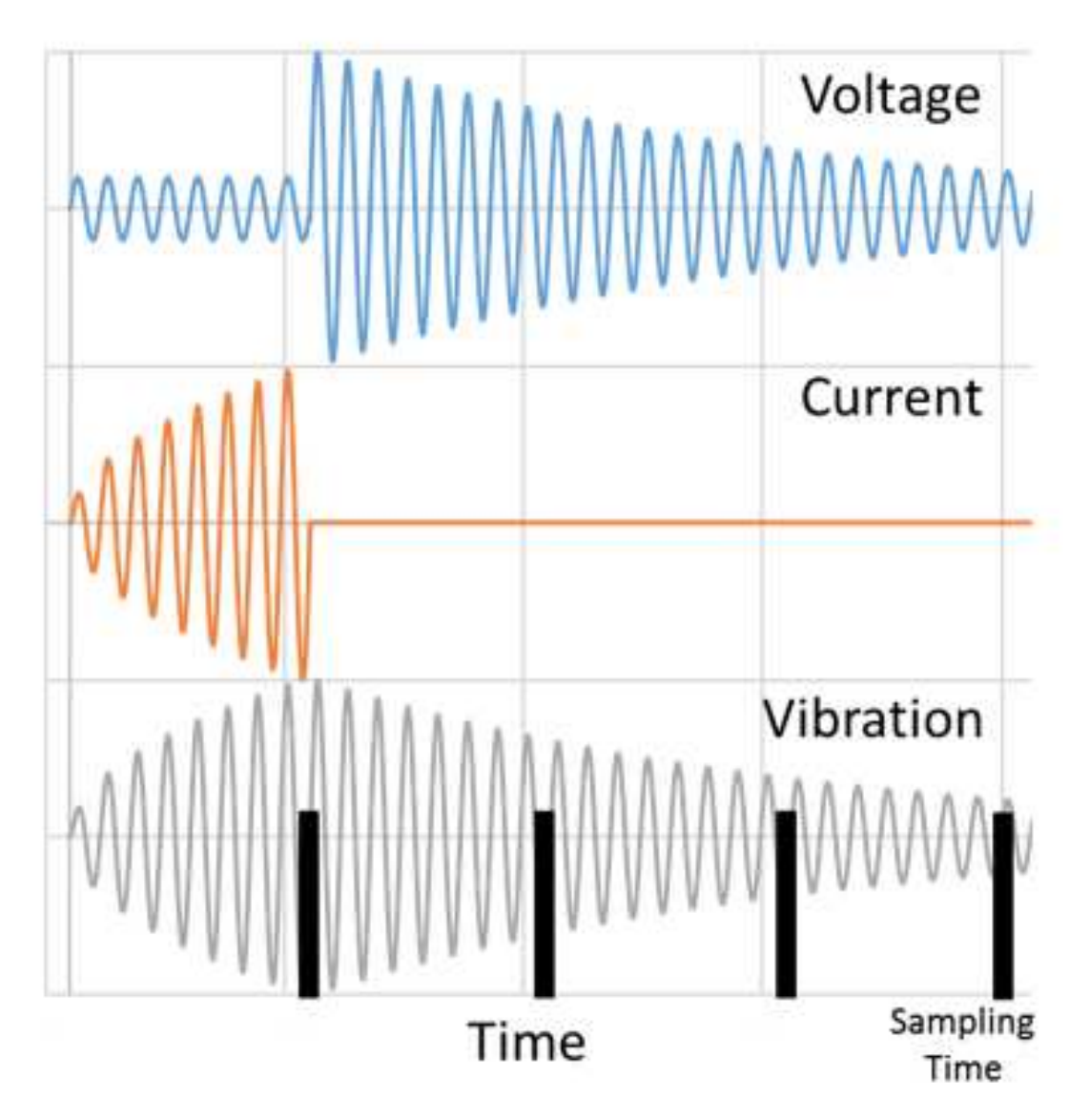}
\begin{description}
\item [{Figure~4}] Qualitative waveforms describing voltage, current,
and vibration (a) before and after short circuit and (b) before and
after open circuit\end{description}
\end{minipage}

\begin{minipage}[t]{1\columnwidth}%
\includegraphics[width=10cm]{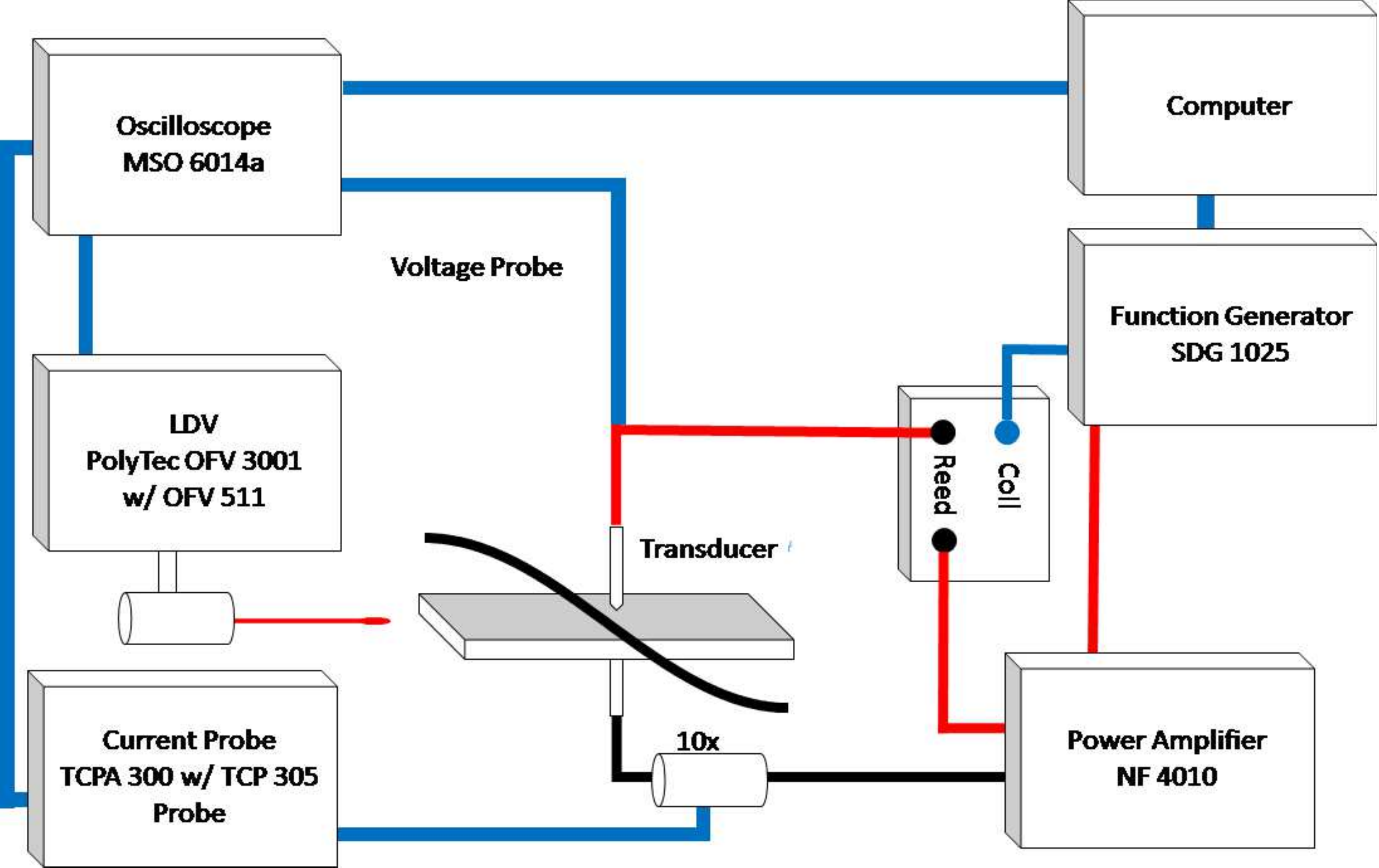}
\begin{description}
\item [{Figure~5}] Experimental setup to apply burst method measurement\end{description}
\end{minipage}

\begin{minipage}[t]{1\columnwidth}%
\includegraphics[width=8cm]{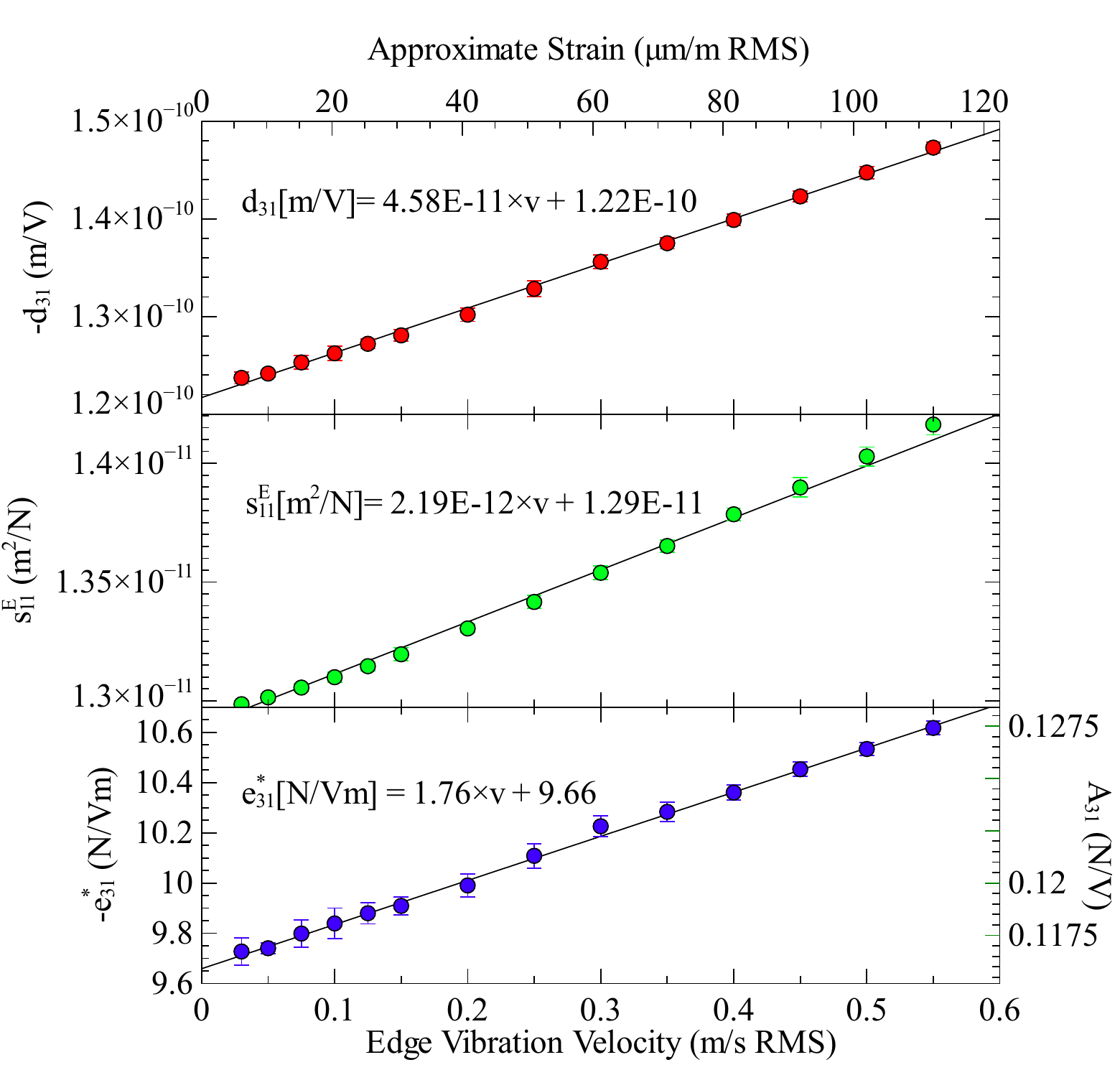}\includegraphics[width=8cm]{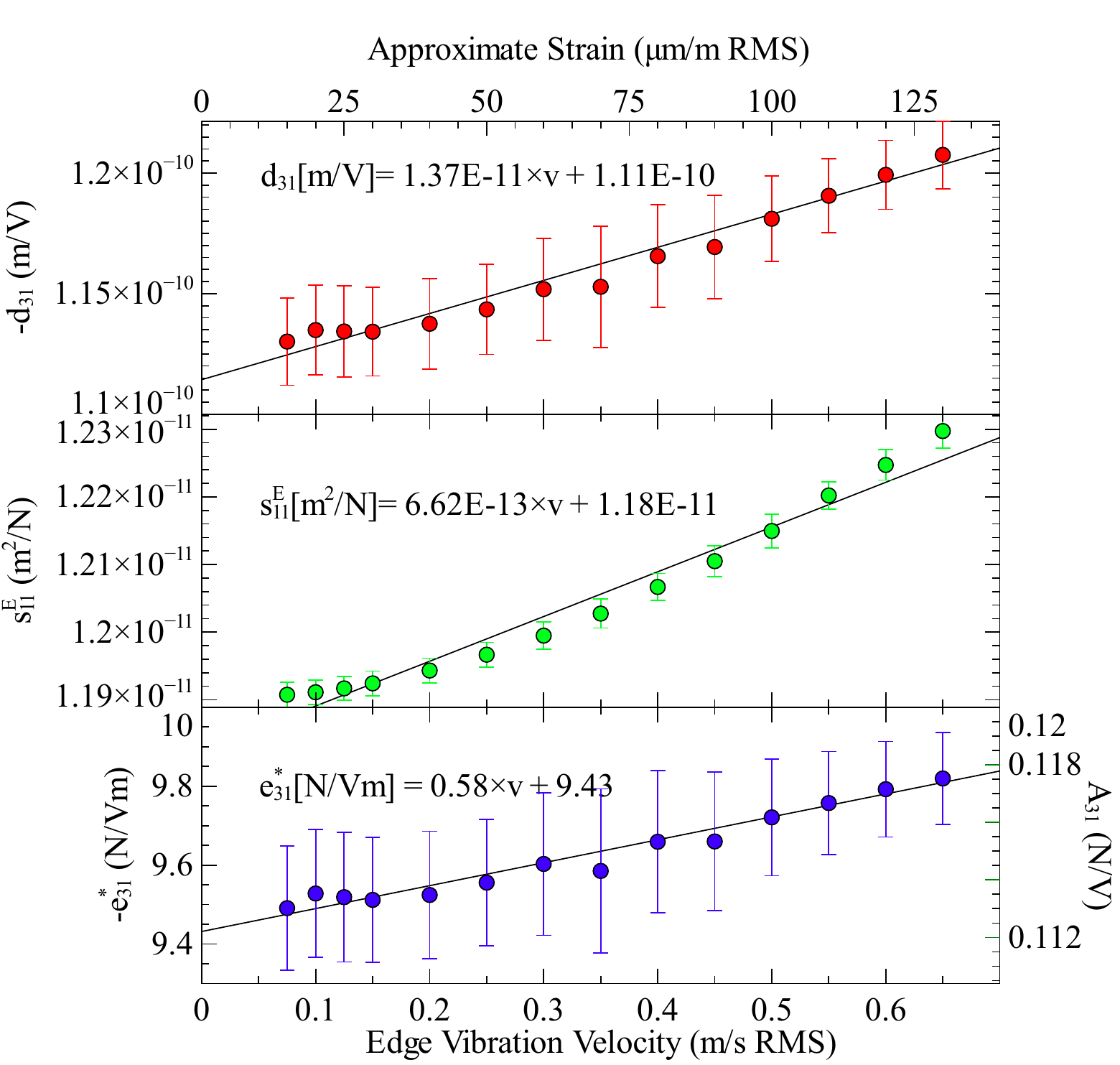}
\begin{description}
\item [{Figure~6}] (a) Resonance characterization of PIC 184 $k_{31}$
and (b) resonance characterization of PIC 144 $k_{31}$\end{description}
\end{minipage}

\begin{minipage}[t]{1\columnwidth}%
\includegraphics[width=8cm]{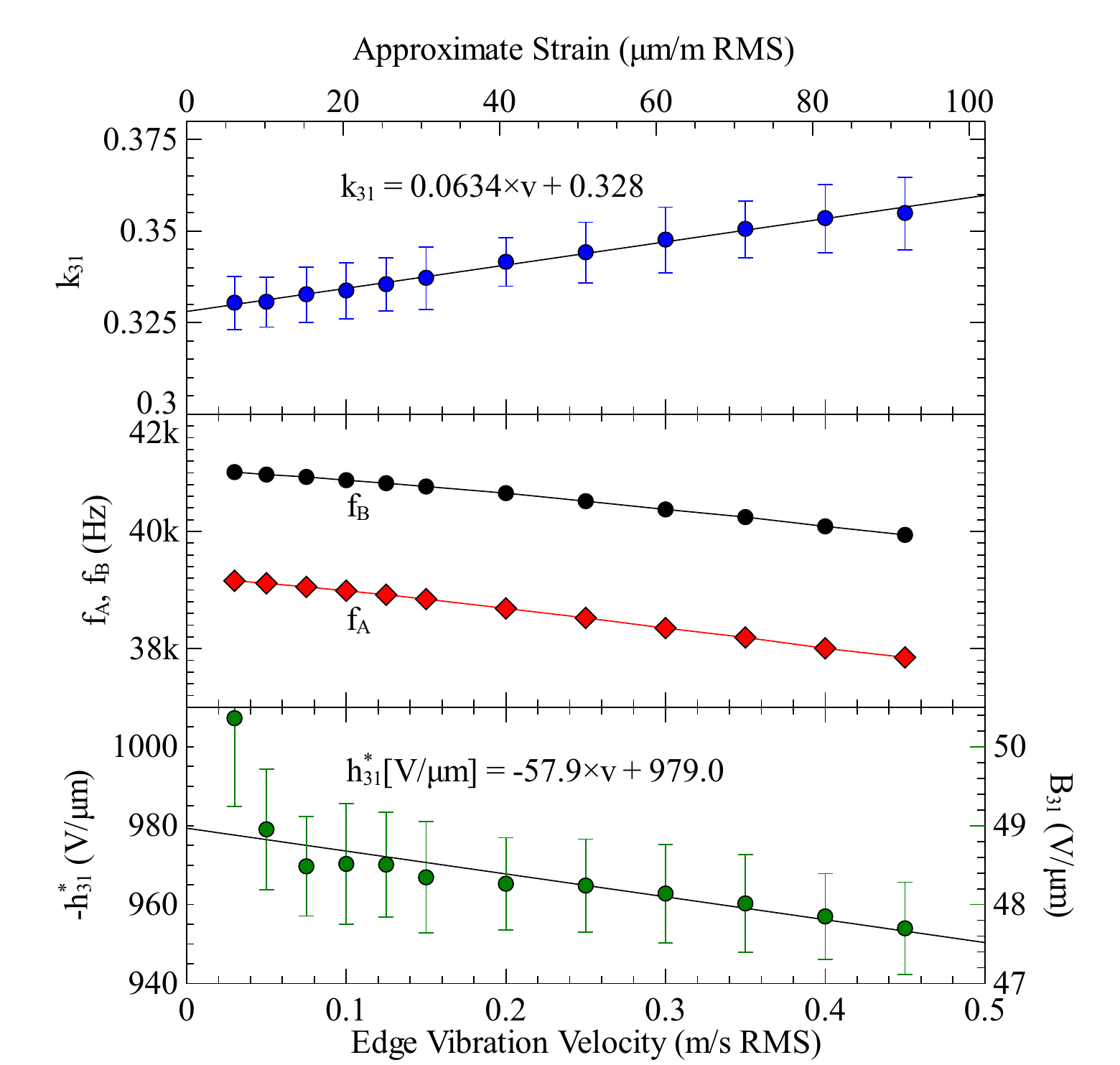}\includegraphics[width=8cm]{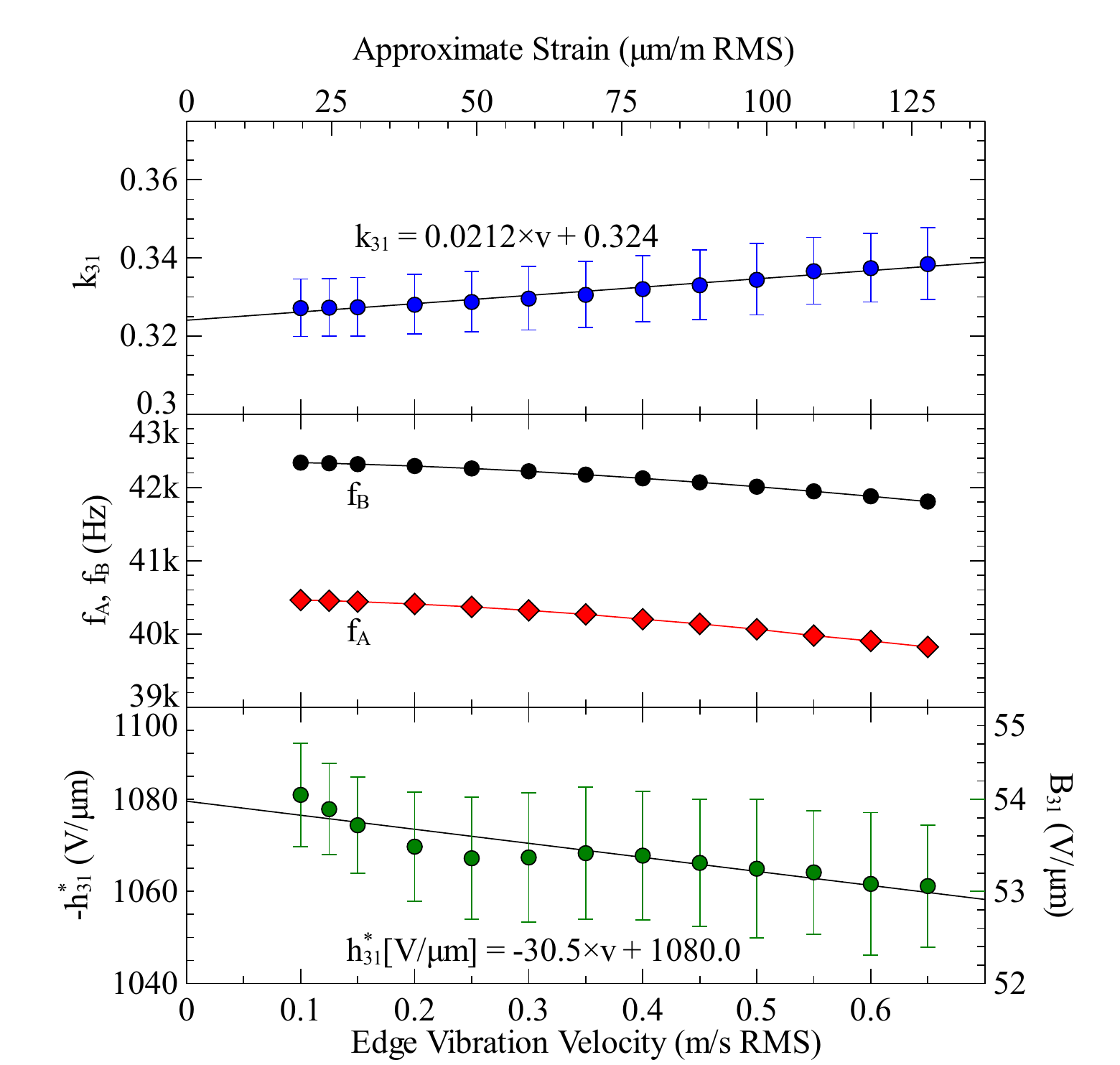}
\begin{description}
\item [{Figure~7}] (a) Antiresonance characterization of PIC 184 $k_{31}$
and (b) antiresonance characterization of PIC 144\end{description}
\end{minipage}

\begin{minipage}[t]{1\columnwidth}%
\begin{description}
\item [{\includegraphics[width=10cm]{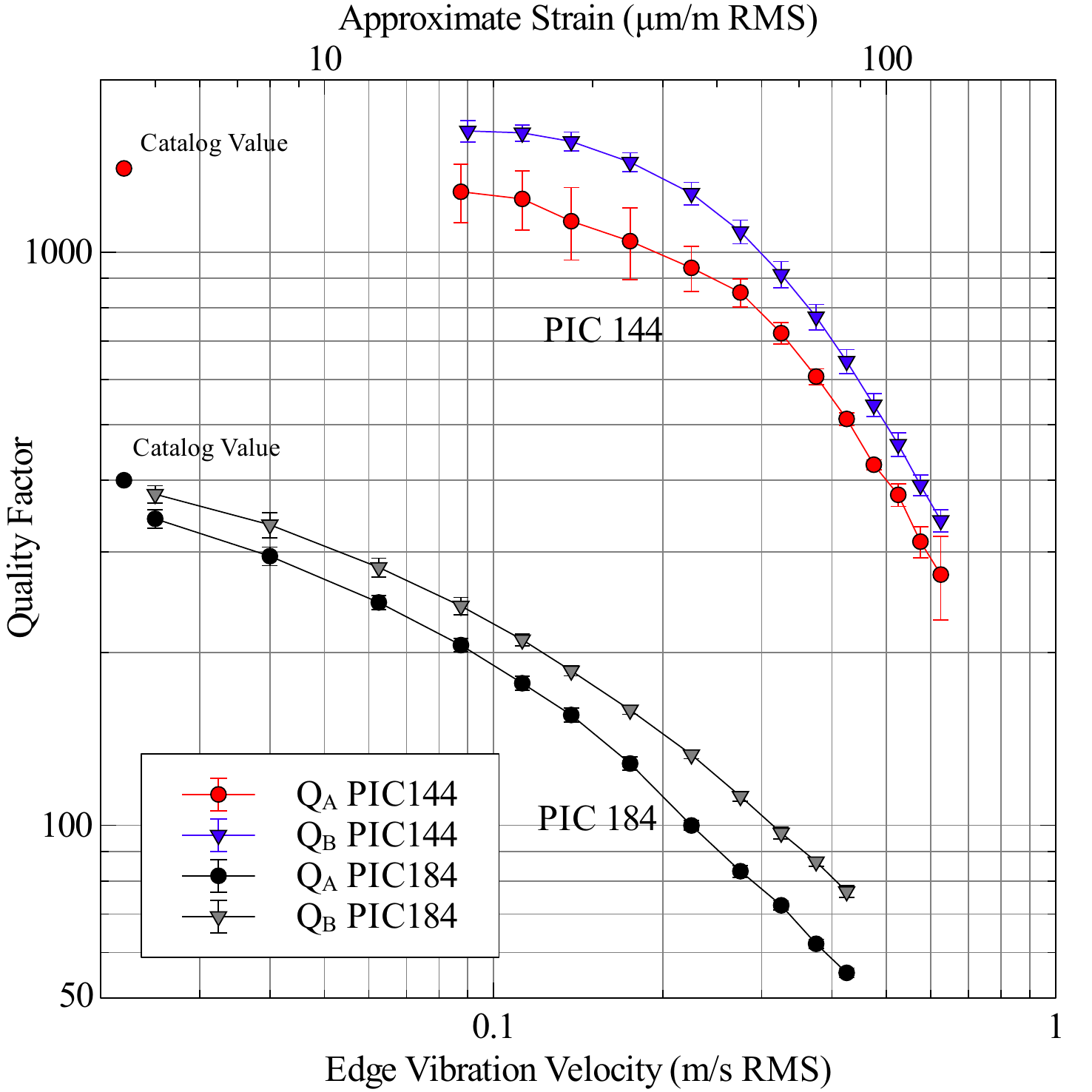}}]~
\item [{Figure~8}] Change in quality factors with vibration velocity for
PIC 184 and PIC 144 $k_{31}$ samples\end{description}
\end{minipage}

\begin{minipage}[t]{1\columnwidth}%
\begin{description}
\item [{\includegraphics[width=10cm]{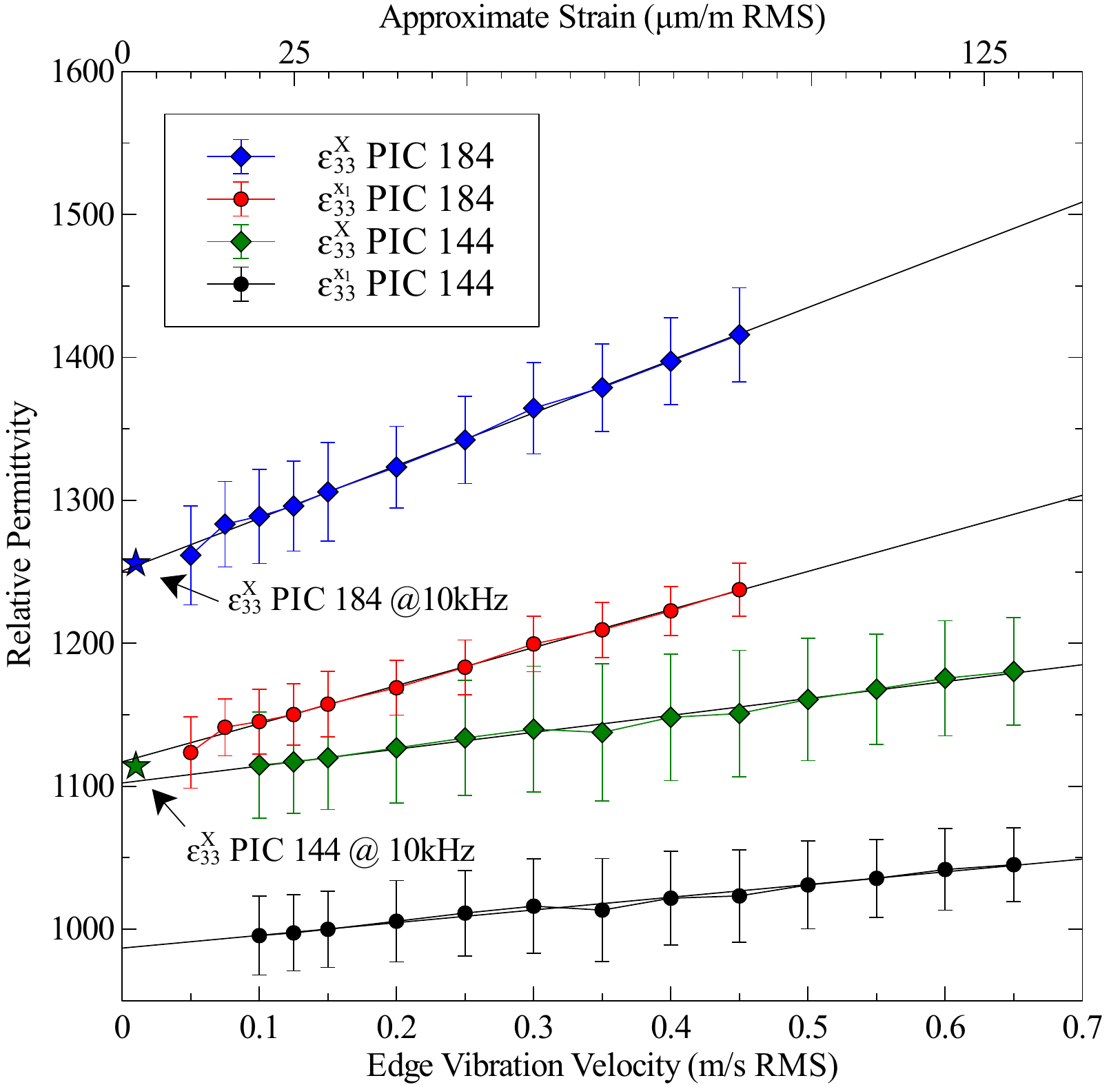}}]~
\item [{Figure~9}] Change in dielectric permittivity with increasing vibration
velocity\end{description}
\end{minipage}

\pagebreak{}

\begin{table}

\section*{Tables}

\begin{centering}
\begin{tabular}{ccc}
Symbol & Description & Unit\tabularnewline
\hline 
$D_{3}$ & electric displacement & C/m$^{2}$\tabularnewline
$E_{3}$ & electric field & V/m\tabularnewline
$\partial u/\partial x$ & strain & -\tabularnewline
$e_{31}^{*}$ & transverse piezoelectric stress constant & N/Vm\tabularnewline
$s_{11}^{E}$ & elastic compliance under constant electric field & m$^{2}$/N\tabularnewline
$d_{31}$ & transverse piezoelectric charge constant & m/V\tabularnewline
$\varepsilon_{33}^{x_{1}}$ & permittivity under constant strain/clamped in the 1-direction & -\tabularnewline
$\varepsilon_{33}^{X}$ & relative permittivity under free stress conditions & -\tabularnewline
$\varepsilon_{0}$ & permittivity of free space & F/m\tabularnewline
$u(x,t)$ & spatial displacement distribution & m\tabularnewline
$t$ & time & s\tabularnewline
$A_{e}$ & area of the electrode & m$^{2}$\tabularnewline
$i(t)$ & current & A\tabularnewline
$i_{0}$ & current amplitude & A\tabularnewline
$\omega$ & angular frequency & rad/s\tabularnewline
$x$ & position along the length (1-direction) & m\tabularnewline
$u_{0}$ & edge displacement amplitude & m\tabularnewline
$v(x,t)$ & spatial vibration velocity distribution & m/s\tabularnewline
$v_{0}$ & edge vibration velocity amplitude & m/s\tabularnewline
\end{tabular}
\par\end{centering}

\raggedright{}\caption{Description of symbols\label{tab:Description-of-symbols}}
\end{table}

\begin{table}[h]
\begin{centering}
\begin{tabular}{ccccccc}
\hline 
 & $\varepsilon_{33}^{X}$@10kHz & $s_{11}^{E}$(m$^{2}$/N)  & $-d_{31}$(m/V) & $k_{31}$ & $\tan\delta'$@10kHz & $Q_{m}$\tabularnewline
\hline 
PIC 184 & 1260 & 1.30E-11 & 124E-12 & 0.330 & 0.004 & 340\tabularnewline
PIC 144 & 1110 & 1.19E-11 & 113E-12 & 0.327 & 0.0015 & 1300\tabularnewline
\hline 
\end{tabular}
\par\end{centering}

\caption{Low power properties of PIC 184 and PIC 144 as calculated from the
burst mode\label{tab:Low-power-properties}}
\end{table}

\end{document}